\documentclass[a4paper,11pt]{article}
\usepackage{pos}
\usepackage{subcaption}
\usepackage{amsmath}
\usepackage{physics}

\newcommand{\rmd}{\mathrm{d}}
\newcommand{\barq}{{\bar{q}}}
\newcommand{\Cf}{C_\mathrm{F}}
\newcommand{\litwo}{\rm Li_2}

\title{Towards mixed QCD$\times$EW corrections to the charged-current Drell-Yan process with leptonic decay at high transverse masses}
\ShortTitle{Mixed QCD$\times$EW corrections to the CCDY process}

\author[a]{Christian Biello}
\author[b]{Amlan Chakraborty}
\author[a]{Giulio Gambuti}
\author[b,c]{Raoul R{\"o}ntsch}
\author*[d]{Chiara Signorile-Signorile}

\def\TIF{Tif Lab, Dipartimento di Fisica, Universit\'{a} di Milano, Via Celoria 16, I-20133 Milano, Italy}
\def\INFNMi{INFN, Sezione di Milano, Via Celoria 16, I-20133 Milano, Italy}
\def\ETH{Institute for Theoretical Physics, ETH Zürich, Wolfgang-Pauli-Str. 27,
8093 Zürich}
\def\CERN{Theoretical Physics Department, CERN, 1211 Geneva 23, Switzerland}

\emailAdd{cbiello@phys.ethz.ch}
\emailAdd{amlan.chakraborty@unimi.it}
\emailAdd{ggambuti@phys.ethz.ch}
\emailAdd{raoul.rontsch@unimi.it}
\emailAdd{chiara.signorile-signorile@cern.ch}

\affiliation[a]{\ETH}
\affiliation[b]{\TIF}
\affiliation[c]{\INFNMi}
\affiliation[d]{\CERN}

\abstract{The Drell–Yan process is a key precision benchmark at the Large Hadron Collider and a sensitive probe of SM and SMEFT effects. Fully exploiting future LHC data requires percent-level SM predictions, motivating precise EW and mixed QCD$\times$EW corrections. In this proceeding we report of the progress towards the calculation of mixed QCD$\times$EW corrections to charged-current Drell-Yan with leptonic decay at high transverse masses, performed within the Nested Soft-Collinear Subtraction framework. After presenting a validation at NLO, we show results for the NNLO real corrections and compare their impact in the neutral-current and newly developed charged-current channels. These contributions are essential to match the current and future precision of LHC measurements, especially in regions where electroweak effects are enhanced by Sudakov logarithms.
\\
\bigskip

CERN-TH-2026-149\\
TIF-UNIMI-2026-8}

\FullConference{Loops and Legs in Quantum Field Theory (LL2026)\\
12-17, April, 2026\\
Bayreuth, Germany\\}

\begin{document}
\maketitle

\section{Introduction}

The Drell--Yan process is one of the most important reactions at hadron colliders, providing a clean probe of the Standard Model and playing a central role in precision measurements and searches for new physics. In particular, charged-current Drell--Yan (CCDY) production, involving a $W$ boson decaying leptonically, is a cornerstone of the LHC electroweak physics program. As the LHC enters the precision era, with experimental uncertainties reaching the percent level, increasingly accurate theoretical predictions are required. While NLO QCD and electroweak (EW) corrections and NNLO QCD calculations are well established for DY-like processes, mixed QCD$\times$EW corrections are becoming essential to match the experimental precision.

This report presents the first steps toward the computation of mixed QCD$\times$EW corrections to the CCDY process, focusing on the high-transverse-mass region. In this regime, electroweak Sudakov logarithms enhance the virtual corrections~\cite{Kuhn:1999nn,Denner:2001gw}, leading to sizeable effects in the tails of differential distributions.

\section{Set up of the calculation}
We focus on the LO process $
pp \to \ell^\pm \nu$
and write schematically the perturbative expansion of the partonic cross section  as
\begin{equation}
\hat{\sigma} = \hat{\sigma}^{\text{LO}} \left(1 + \alpha_s  \, \delta \sigma^{\text{NLO}}_{\text{qcd}} 
+ \alpha_{\text{ew}} \, \delta \sigma^{\text{NLO}}_{\text{ew}} 
+ \alpha_s^2 \, \delta \sigma^{\text{NNLO}}_{\text{qcd}} 
+ \alpha_s \alpha_{\text{ew}} \, \delta \sigma^{\text{NNLO}}_{\text{mix}} 
\, + \dots \right) \, 
\end{equation}
where $\alpha_s$ and $\alpha_{\text{ew}}$ are the strong and electroweak coupling constants respectively. 
From a power-counting perspective, NLO QCD corrections are typically the dominant effect, while NLO electroweak and NNLO QCD contributions are at the percent level. Although mixed QCD$\times$EW corrections are formally suppressed, electroweak Sudakov logarithms can enhance them significantly in high-energy regions, making them relevant for precision predictions.

The computation of mixed QCD$\times$EW corrections faces two main challenges: the evaluation of two-loop amplitudes and the treatment of real radiation. The virtual contributions involve multi-scale two-loop Feynman integrals, which have been addressed through both analytic and semi-analytic approaches (see for instance Ref.~\cite{Heller:2019gkq, Heller:2020owb, Armadillo:2022bgm} for neutral-current calculations, and Ref.~\cite{Armadillo:2024nwk} for charged current results). Real radiation contributions generate infrared singularities from soft and collinear emissions, which are handled using the nested soft-collinear subtraction (NSC) method~\cite{Caola:2017dug}. Its extension to mixed corrections must consistently account for both gluon and photon radiation.

\section{NLO Electroweak Corrections and validation against \texttt{POWHEG}}

\begin{figure}[h!]
\begin{center}
\begin{tabular}{ccc}
\hspace{-1.0cm}
\includegraphics[width=.36\textwidth, page=1]{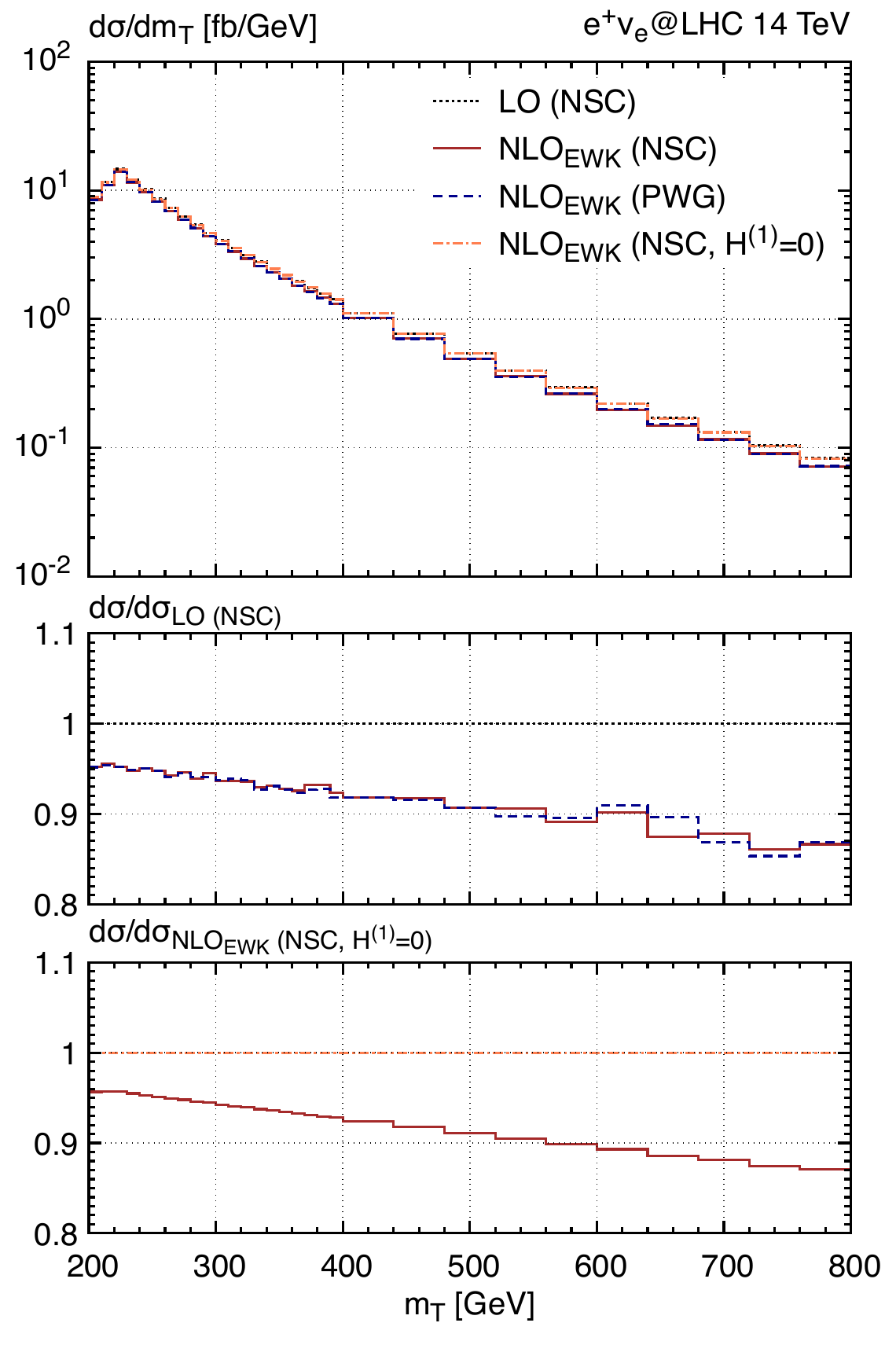}&\hspace{-0.5cm}
\includegraphics[width=.36\textwidth, page=2]{figures/allplots-NLOEW-Wp.pdf}&\hspace{-0.5cm}
\includegraphics[width=.36\textwidth, page=3]{figures/allplots-NLOEW-Wp.pdf}
\end{tabular}
\vspace*{1ex}
\caption{Differential distributions for charged-current Drell--Yan production,
$pp\to e^+\nu_e+X$, at the LHC with $\sqrt{s}=14~\mathrm{TeV}$.
The upper panels show the $m_{\rm T}$, $m_{e^+\nu_e}$, and $p_{{\rm T},e^+}$ spectra
at LO and NLO EW accuracy. The middle panels display the ratios to the LO
prediction, while the lower panels compare the full NLO EW result with the
approximation obtained by setting $H^{(1)}=0$. Results from the NSC and \texttt{POWHEG} (PWG)
implementations are shown for comparison. See the text for details.}
\label{fig:WpNLOEW}
\end{center}
\end{figure}

As a first step, the NLO QCD and the NLO electroweak corrections have been computed within the NSC framework. 
These corrections provide a useful benchmark and allow for a detailed study of the structure of the calculation.
At the integrated level, the NLO electroweak corrections to the CC-DY process amount to approximately $-3\%$ of the leading-order cross section. However, their impact is significantly larger at the differential level. In particular, in the transverse mass range between $200$ and $800$ GeV, the corrections can reach values of order $10\%$, reflecting the presence of Sudakov logarithms.
To validate the calculation, we developed an independent implementation in the \texttt{POWHEG} framework~\cite{Jezo:2015aia}. The comparison shows excellent agreement and provides strong evidence for the correctness of the subtraction procedure and the numerical implementation.
In Fig.~\ref{fig:WpNLOEW} we present a selection of the observable investigated for the $W^+$-mediated di-lepton process at NLO EW: the transverse mass $m_{\rm T}$, the reconstructed
invariant mass $m_{e^+\nu_e}$, and the positron transverse momentum
$p_{{\rm T},e^+}$.
In all observables, the NLO EW corrections reduce the LO prediction over the entire kinematic range. Their magnitude increases with the characteristic energy scale of the process, reaching approximately $10\%$--$13\%$ in the highest bins considered. This behaviour is consistent with the expected
enhancement of negative electroweak Sudakov logarithms at large momentum
transfers. The
differences between the results obtained within the NSC and PWG frameworks remain at the sub-percent level throughout the studied phase
space, confirming the consistency of the two implementations. The lower ratio
panels further illustrate the impact of the finite-virtual contribution $H^{(1)}$ in the minimal-subtraction scheme: when this term is omitted, the resulting prediction differs from
the full NLO EW result by up to about $10\%$, demonstrating that $H^{(1)}$
provides a sizeable and non-negligible contribution to the electroweak
corrections across the considered kinematic range.

\section{Organisation of the results for mixed corrections at NNLO}

After the subtraction of infrared singularities, the remaining finite contributions can be organized into different components corresponding to double-virtual, real-virtual, and double-real contributions. These are further decomposed into fully-resolved (FR), single-unresolved (SU), and double-unresolved (DU) contributions, which we specify below.   In terms of these three contributions the result reads 
\begin{equation}
2s_{ab} \, \rmd \hat{\sigma}_{ab}^{\rm NNLO}
= 2s_{ab} \big[\rmd \hat{\sigma}_{ab}^{\rm FR} 
+ \rmd \hat{\sigma}_{ab}^{\rm SU} 
+ \rmd \hat{\sigma}_{ab}^{\rm DU} 
\big] \,.
\label{eq_final_result_dec}
\end{equation} 
The fully resolved contributions correspond to four-body kinematics and are evaluated numerically. The phase space is divided into sectors, each treated independently, allowing for efficient parallelization. In each sector, only a limited number of kinematic configurations need to be considered, simplifying the numerical integration.
The resulting expressions are fully differential and locally finite, leading to stable numerical evaluations. Although the analytic expressions can be quite involved in a general reference frame, they simplify considerably in the partonic centre-of-mass frame, where compact structures emerge. For the sake of illustration, we focus on the partonic channel $q(p_1)+\barq'(p_2) \rightarrow \ell(p_3) +\bar{\ell}'(p_4)(+ g(p_5)+\gamma(p_6))$, define the Born-level configuration as     $\mathcal{H} = 1_q, 2_{\barq'}, 3_{\ell}, 4_{\bar{\ell}'}$. The fully-resolved contribution that can be directly implemented in four dimensions in a Monte Carlo framework reads~\footnote{Notation and conventions are the same as in Ref.~\cite{Buccioni:2022kgy} and ~\cite{Devoto:2025jql}.}
\begin{equation}
\rmd \hat{\sigma}^{\rm FR}_{q\bar{q}'}
=
\big\langle (I-S_g) (I-S_\gamma) \, \Omega^{\, q\bar{q}'}_1 F_{\rm LM}(\mathcal{H}|5_g,6_\gamma)\big\rangle \; ,
\end{equation}
with 
\begin{equation}
\label{eq:Omega1}
\begin{split} 
\Omega^{\, q\bar{q}'}_1 & = 
\sum_{i=1,2}
(1-C_{g\gamma, i}) 
\Big[
  (1-C_{g i})  \, \omega^{\gamma i,g i}  \, \theta_a^{(i )} 
 +   (1-C_{\gamma i}) \, \omega^{\gamma i,g i } \, \theta_c^{(i)} 
 \Big]
 \\
&
+
\sum_{\substack{i,j = 1, 2 \\ i \neq j }}
(1-C_{g j })  (1-C_{\gamma i }) \, \omega^{\gamma i,g j}
+ \sum_{i=3,4}\sum_{j=1,2} (1-C_{gj})  (1-C_{\gamma i})  \, \big[ \omega^{\gamma i,gj} 
+ \, \omega^{\gamma i,gj} \big] \; .
\end{split}
\end{equation}
In the equation above, $S_i$ represent the single soft operator that extract the leading behaviour in the $E_i \rightarrow 0$ limit. $C_{kl}$ (related to the $p_k \parallel p_l$ limit) and $C_{kl,i}$ (related to the $p_k \parallel p_l \parallel p_i$ limit) identifies the double and triple collinear operator, respectively. The $\omega^{\gamma i,gi}$ and $\omega^{\gamma i,gj}$ are the triple-collinear and double-collinear partition functions, which each isolate a well-defined subset of collinear singularities. Triple-collinear partitions further require the introduction of sectors ($1= \theta^{(a)}+\theta^{(c)}$) to 
disentangle strongly-ordered configurations. Upon fixing the unresolved partons to be $5_g, 6_\gamma$, only two distinct sectors are required to implement the fully-resolved contributions for
each of the 2 triple-collinear partitions, namely $ \theta^{(\alpha)}\omega^{\gamma i, g i} \equiv \theta^{(\alpha)}\omega^{g i, \gamma i} = \{\theta^{(a)} \, \omega^{51, 61}, \theta^{(a)} \, \omega^{5 2, 62}, 
\theta^{(c)} \, \omega^{51, 61}, \theta^{(c)} \, \omega^{5 2, 62}\}$. 
The double-collinear sectors required for this calculation are selected by the functions $\omega^{51,62}, \omega^{51,63}$, $\omega^{51,64}$, $\omega^{52,61}$, $\omega^{52,63}$, $\omega^{52,64}$.
We note that each 
sector can be parameterized 
and run independently~\footnote{In the rest of the manuscript we will refer to the sectors as ($5161a$, $5161c$, $5262a$, $5262c$,  
$5162$, $5163$,
$5164$, $5261$, 
$5263$, $5264$).}. In Sec.~\ref{sec:doubleunresolved} we present the numerical impact of the various sectors and partitions, and compare the results with the NCDY predictions.

Next, we consider the single-unresolved contribution, which can be decomposed into terms with boosted and un-boosted kinematics. The subset of the former involving only the first initial state reads
\begin{equation}
\begin{split}
\rmd \hat{\sigma}^{{\rm SU, \,  sb, \, 1}}_{q\bar{q}'}
={}&
[\alpha_{\rm ew}] \, 
 \Big\langle
 Q_q^2 \; 
  \mathcal{O}_{\rm nlo}^g \,  
  \Big[
  \tilde{P}_{qq}^{\rm NLO}(E_c, z)
  + \tilde{\omega}_{\gamma \parallel 1}^{\gamma 1, g 1} \, 
  \log \eta_{15} \; \bar{P}_{qq, R}^{\rm AP, 0}(z)
  \Big] \otimes
 F_{\rm LM}(\mathcal{H}| 5_g) 
 \Big\rangle
 \\
 +{}&
 [\alpha_s] \, 
 \Big\langle
 \Cf \, 
  \mathcal{O}_{\rm nlo}^\gamma \, 
  \Big[
  \tilde{P}_{qq}^{\rm NLO}(z) 
  + \tilde{\omega}_{g\parallel 1}^{\gamma 1, g 1} \, 
  \log \eta_{15} \; \bar{P}_{qq, R}^{\rm AP, 0}(z)
  \Big] \otimes
 F_{\rm LM}(\mathcal{H}| 5_\gamma) \Big\rangle \,,
\end{split}
\end{equation}
where the $\mathcal{O}_{\rm nlo}^\alpha$ is the fully resolved operator defined as $\mathcal{O}_{\rm nlo}^\alpha = (1-S_{\alpha})(1-\sum_{i} C_{\alpha i})$. The functions $\tilde{P}_{qq}^{\rm NLO} (z, E)$ and $\bar{P}_{qq, R}^{\rm AP, 0}(z)$ are related to the tree-level splitting functions~\footnote{The explicit definition can be found in Eq.~(2.66) in Ref.~\cite{Buccioni:2022kgy}} and the symbol $\otimes$ indicate the convolutions of such splitting functions with the matrix elements where the incoming momentum $p_1$ is rescaled by a the energy fraction factor $z$. Next, the $\tilde{\omega}_{\alpha\parallel 1}^{\gamma 1, g 1}$ indicates the result of applying the limit $\alpha\parallel 1$ onto the partition function $\omega^{\gamma 1, g 1}$. Finally we have $\eta_{ab} = (1-\cos \theta_{ab})/2$, with $\theta_{ab}$ being the angle between the directions of partons $a$ and $b$. The elastic (un-boosted) component of the SU contribution is given by
\begin{equation}
\begin{split}
\rmd \hat{\sigma}^{{\rm SU, \,  el}}_{q\bar{q}'}
= {}&
\big\langle
\mathcal{O}_{\rm nlo}^g \, 
 F_{\rm RV\,, fin}^{\rm (EW)} (\mathcal{H}| 5_g)
 \big\rangle
 +
\big\langle
 \mathcal{O}_{\rm nlo}^\gamma \, 
 F_{\rm RV, \, fin}^{\rm (QCD)} (\mathcal{H}| 5_\gamma)
 \big\rangle
\\
{}&  
+  [\alpha_{\rm ew}] \,  \Big\langle
 \mathcal{O}_{\rm nlo}^g \, 
 \Big[ \mathcal{G}_{\rm nlo}^{\rm (EW)}
 F_{\rm LM} (\mathcal{H}| 5_g)
 \Big]
\Big \rangle  
+  [\alpha_{\rm s}] \,  \Big\langle
 \mathcal{O}_{\rm nlo}^\gamma \, 
 \Big[\Cf\,  \mathcal{G}_{\rm nlo}^{\rm (QCD)}
 F_{\rm LM} (\mathcal{H}| 5_\gamma)
 \Big]
\Big \rangle  \; ,
\end{split}
\end{equation}
where in the first line we have introduced the finite remainder of the real-virtual corrections, $F_{\rm RV\,, fin}^{\rm (EW)}$ ($F_{\rm RV, \, fin}^{\rm (QCD)}$), involving one EW (QCD) virtual exchange and one QCD (EW) emission. In the second line, we have introduced the functions
\begin{align}
\mathcal{G}_{\rm nlo}^{\rm (QCD)}
 \;=\;& \frac{2 \pi^2}3 + 6 L_c + 3 \log \eta_{12} + 2 \litwo(1-\eta_{12})
\\
 \mathcal{G}_{\rm nlo}^{\rm (EW)}
 \;=\;&
Q_\ell^2\,A_\ell
+ Q_\ell Q_q\,A_{\ell q}(\eta_{13})
+ Q_\ell Q_{\barq'}\,A_{\ell q}(\eta_{23})
+ Q_q Q_{\barq'}\,A_{qq}(\eta_{12})
\end{align}
which involve the logarithmic term $L_c=\log\left(2 E_c/\mu\right)$ where $E_c$ is the partonic center-of-mass energy. The coefficient functions $A$ are simple combinations of constants, $L_c$, $L_\ell= \log\left(2 E_\ell/\mu\right)$, $\log(\eta)$, where $E_\ell$ is the energy of the charged lepton, and $\litwo(1-\eta)$.
In the next section we present results for a selection of the contributions mentioned above, emphasising the differences with respect to the NCDY production.

\section{Selection of contributions and comparison with neutral-current Drell-Yan}
\label{sec:plots}

\paragraph{Double unresolved corrections.} \label{sec:doubleunresolved}
We then turn to discussing the implementation of the fully resolved contribution, focusing on the fully resolved component. We present the 
comparison of the equivalent correction for both NCDY and CCDY to 
emphasise some relevant differences and similarities. 
\begin{table}[t]
\centering
\small
\begin{tabular}{lc|lc}
\hline
\multicolumn{4}{c}{\textbf{Double-real sectors contributions: NCDY vs CCDY ($W^+$)}} \\
\multicolumn{4}{c}{$(100~\text{seeds},\ 5.0\times10^{6}\times15\ +\ 5.0\times10^{7}\times10)$} \\
\hline
\multicolumn{2}{c|}{NCDY} & \multicolumn{2}{c}{CCDY ($W^+$)} \\
{\it Sector} & {\it Cross-section} [fb] & {\it Sector} & {\it Cross-section} [fb] \\
\hline
5161a & $0.93220 \pm 0.00011$ & 5161a & $0.17887 \pm 0.00009$ \\
5262a & $0.93222 \pm 0.00011$ & 5262a & $0.17873 \pm 0.00009$ \\
\hline
5161c & $0.04374 \pm 0.00030$ & 5161c & $0.04200 \pm 0.00017$ \\
5262c & $0.04356 \pm 0.00031$ & 5262c & $0.04227 \pm 0.00017$ \\
\hline
5163 & $0.08572 \pm 0.00024$ & 5163 & $0.05167 \pm 0.00005$ \\
5263 & $0.08647 \pm 0.00025$ & 5263 & $0.05177 \pm 0.00005$ \\
\hline
5164 & $0.12337 \pm 0.00036$ & 5164 & $0.12578 \pm 0.00038$ \\
5264 & $0.12344 \pm 0.00036$ & 5264 & $0.12549 \pm 0.00038$ \\
\hline
5162 & $0.36610 \pm 0.00023$ & 5162 & $0.22284 \pm 0.00013$ \\
5261 & $0.36571 \pm 0.00020$ & 5261 & $0.22296 \pm 0.00013$ \\
\hline
\end{tabular}
\caption{Comparison of representative double-real subtraction sectors in NCDY and CCDY ($W^+$). The close agreement within the pairs $(5161a,5262a)$, $(5161c,5262c)$, $(5163,5263)$, $(5164,5264)$, and $(5162,5261)$ reflects the sector symmetries discussed in the text. The results are produced by combining 100 independent seeds, each obtained with 15 integration iterations to pre-sample the phase space with $5 \times 10^6$ points, and then 10 iterations with $5 \times 10^7$ points.}
\label{tab:sector_comparison_ncdy_ccdy}
\end{table}
Table~\ref{tab:sector_comparison_ncdy_ccdy} reports the integrated
contributions of representative double-real subtraction sectors for both
NCDY and CCDY ($W^+$). In agreement with the expectations from the
singularity structure of the subtraction framework, sectors naturally group
into pairs with nearly identical integrated cross sections. This pairing is
particularly evident for the $(5161a,5262a)$, $(5161c,5262c)$,
$(5163,5263)$, $(5164,5264)$ and $(5162,5261)$ sectors, whose results agree
within the numerical integration uncertainties. The existence of these
sector pairs reflects the fact that they are associated with analogous
singular limits. While this pattern is present in both neutral- and charged-current Drell--Yan production, the hierarchy among sector pairs differs significantly. In NCDY, the contributions are strongly ordered, with some pairs suppressed by more than an order of magnitude relative to the dominant ones. In CCDY, the spread is considerably reduced, resulting in a more uniform distribution of the integrated cross section across the different singular configurations.
\begin{figure}[h!]
\begin{center}
\begin{tabular}{cc}
\hspace{-2.0cm}
\includegraphics[width=0.60\textwidth]{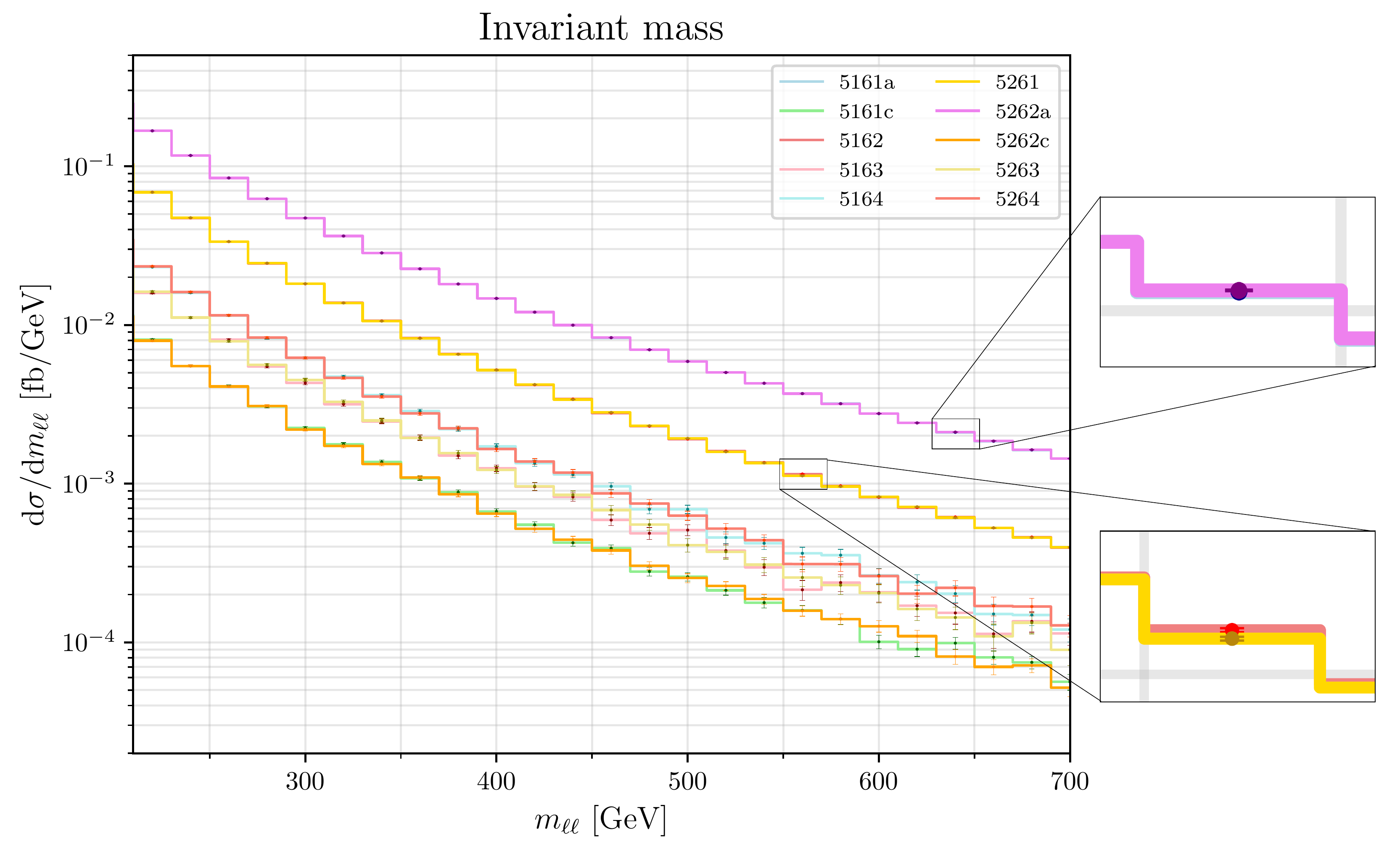}
&
\includegraphics[width=0.62\textwidth]{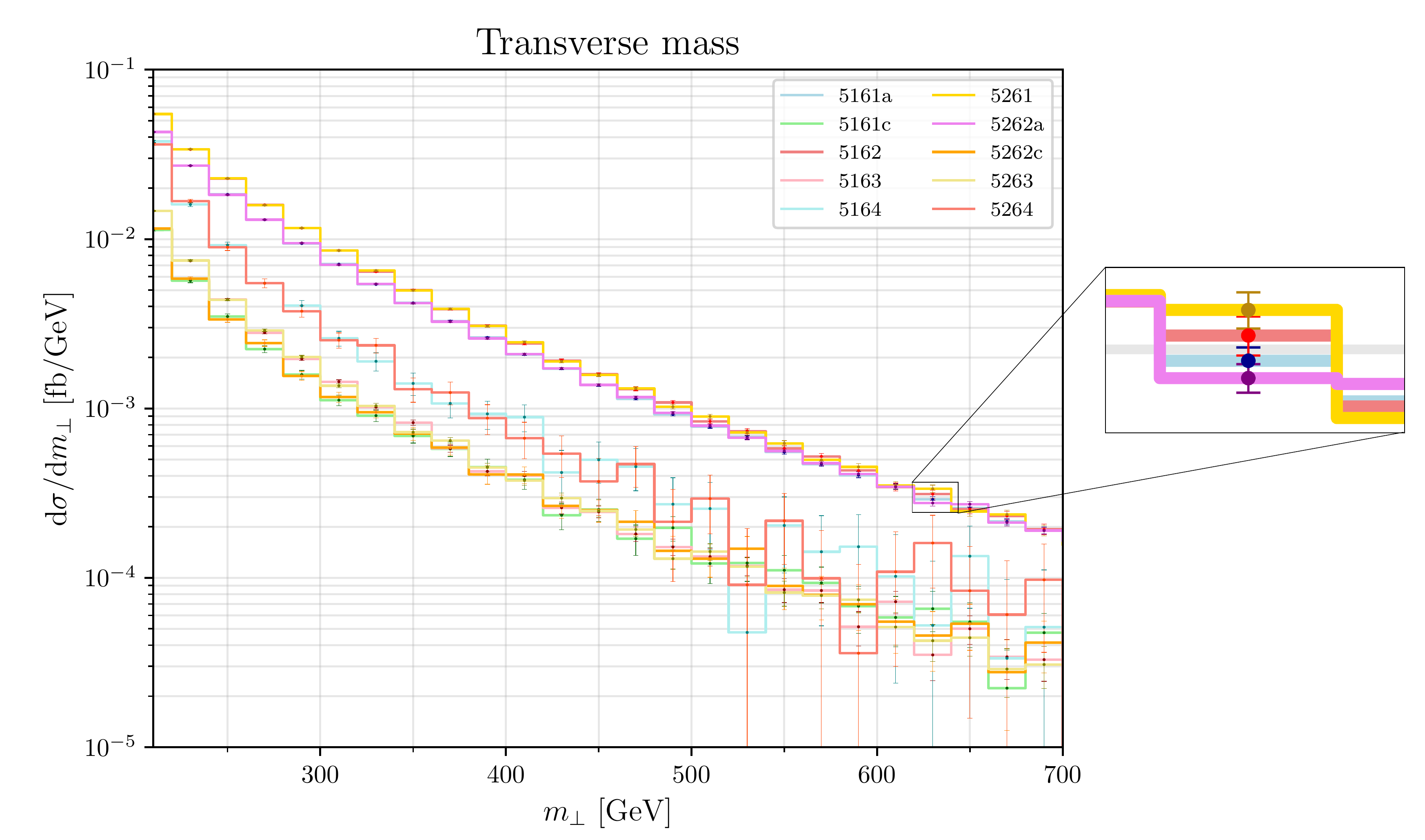}
\end{tabular}
\vspace*{1ex}
\caption{Comparison of the contribution of different double-real sectors to a given distribution: 
dilepton-invariant mass distribution for the neutral-current Drell--Yan process (left),
and W-boson transverse mass distribution for the charged-current Drell--Yan process (right).}
\label{fig:DY_RR_sectors}
\end{center}
\end{figure}

Figure~\ref{fig:DY_RR_sectors} presents the sector decomposition of the finite double-real contribution at differential level. The patterns largely mirror those observed for the integrated cross section, but are more pronounced in the distributions. In NCDY, the sectors form nearly degenerate pairs across the full kinematic range, while in CCDY this structure is broken, leading to a more homogeneous distribution among sectors. The different relative importance of triple- and double-collinear sectors further highlights the distinct singularity structure of the two processes.

\begin{figure}[h]
    \centering
    \begin{minipage}{0.49\textwidth}
        \centering
        \includegraphics[width=\linewidth]{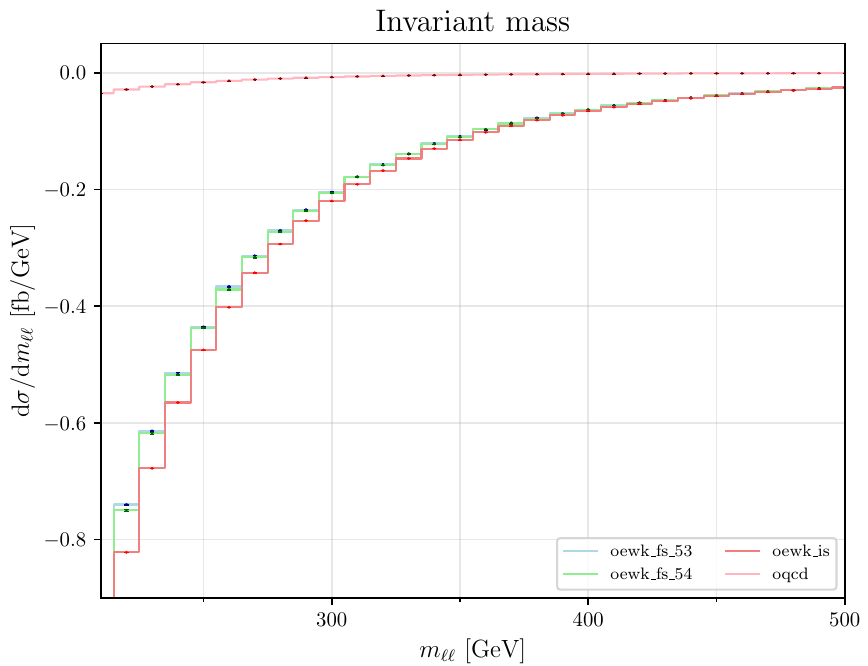}
        \label{fig:fig1}
    \end{minipage}
    \begin{minipage}{0.49\textwidth}
        \centering
        \includegraphics[width=\linewidth]{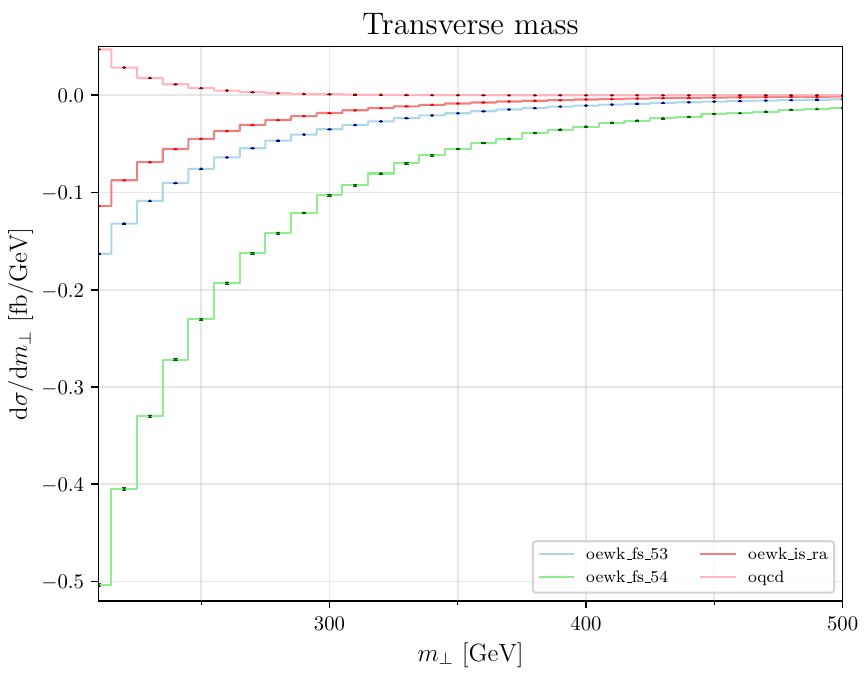}
        \label{fig:fig2}
    \end{minipage}
    \caption{Comparison of
    the contribution of different double-real sectors for neutral-current (top) and charged-current (bottom) Drell-Yan production.}
    \label{fig:singleunresolved}
\end{figure}

\paragraph{Single unresolved corrections.}
As shown in Table~\ref{tab:ncdy_ccdy_single}, the CC process displays a less symmetric sector structure than NCDY,  already at the single-unresolved level, leading to a different hierarchy among the dominant contributions.
Figure~\ref{fig:singleunresolved} compares representative single-unresolved contributions in neutral- and charged-current production. While the NC sectors exhibit a nearly symmetric behaviour, reflecting the underlying process symmetries, the CC channel shows sizeable asymmetries between analogous sectors due to the charged lepton--neutrino final state and the different partonic flavour structure. In addition, the EW contributions are significantly reduced in magnitude in CCDY, leading to smaller cancellations among sectors. These features illustrate how the singularity structure and sector hierarchy become more process dependent in the CC case.

\begin{table}[h!]
\centering
\begin{tabular}{lc|lc}
\hline
\multicolumn{4}{c}{\textbf{$\mathcal{O}_{\rm nlo}$ contributions: NCDY vs CCDY ($W^+$)}} \\
\multicolumn{4}{c}{$(100~\text{seeds},\ 5.0\times10^{5}\times15\ +\ 5.0\times10^{6}\times10)$} \\
\hline
\multicolumn{2}{c|}{NCDY} & \multicolumn{2}{c}{CCDY ($W^+$)} \\
{\it Contribution} & {\it Cross-section} [fb] & {\it Contribution} & {\it Cross-section} [fb] \\
\hline
oqcd        & $-0.27737 \pm 0.00029$ & oqcd        & $0.20859 \pm 0.00008$ \\
\hline
oewk\_fs\_53 & $-7.56029 \pm 0.00214$ & oewk\_fs\_53 & $-1.32523 \pm 0.00015$ \\
oewk\_fs\_54 & $-7.61950 \pm 0.00252$ & oewk\_fs\_54 & $-4.04229 \pm 0.00139$ \\
oewk\_is     & $-8.21967 \pm 0.00069$ & oewk\_is     & $-0.77722 \pm 0.00019$ \\
\hline
rvqcd\_is     & $3.18048 \pm 0.00015$ & rvqcd\_is     & $0.72593 \pm 0.00016$ \\
\hline
\end{tabular}
\caption{Comparison of SU contributions between NCDY and CCDY ($W^+$). The results are produced by combining 100 independent seeds, each obtained with 15 integration iterations to pre-sample the phase space with $5 \times 10^5$ points, and then 10 iterations with $5 \times 10^6$ points.}
\label{tab:ncdy_ccdy_single}
\end{table}

\section{Outlook and Conclusions}

The computation of mixed QCD$\times$EW corrections to the charged-current Drell--Yan process represents a significant step toward achieving the level of precision required for current and future collider experiments. The results obtained so far demonstrate that these corrections, although formally subleading, play an important role in precision phenomenology, especially in the high-energy regime.
Looking ahead, further developments are expected in several directions. These include the completion of the full NNLO calculation, improvements in numerical efficiency. Moreover, the possibility of deriving mixed corrections for arbitrary hadronic scattering through the abelianization of QCD results offers an intriguing avenue for future research.

\section*{Acknowledgments}
\noindent 
C.B. acknowledges funding
from the Swiss National Science Foundation, grant 10001706. C.B.\ and C.S-S. wish to thank the Physics Department of the University of Milan for hospitality during work on this project. The work of R.R. and A.C. is supported by the
Italian Ministry of Universities and Research (MUR) through grant PRIN2022BCXSW9.

\end{document}